\documentstyle[12pt,amssym,aasms4]{article}

\slugcomment{submitted to Ap.J.}

\lefthead{Gambera et al.}
\righthead{A 3-D wavelet analysis of the Coma cluster}

\begin{document}

\title{A  3-D wavelet analysis of substructure in the Coma cluster: \\
statistics and morphology}

\author{M. Gambera}
\affil{Istituto di Astronomia, Universit\`a di Catania, Italy}
\authoraddr{Istituto di Astronomia, Universit\`a di Catania, Citt\`{a} 
Universitaria, Viale A. Doria 6, I-95125 Catania, Italy}
\author{A. Pagliaro}
\affil{Istituto di Astronomia, Universit\`a di Catania, Italy}
\authoraddr{Istituto di Astronomia, Universit\`a di Catania, Citt\`{a} 
Universitaria, Viale A. Doria 6, I-95125 Catania, Italy}
\author{V. Antonuccio-Delogu\altaffilmark{1}}
\affil{Osservatorio Astrofisico di Catania, Italy}
\authoraddr{Osservatorio Astrofisico di Catania, Citt\`{a} 
Universitaria, Viale A. Doria 6, I-95125 Catania, Italy}
\and
\author{U. Becciani}
\affil{Osservatorio Astrofisico di Catania, Italy}
\authoraddr{Osservatorio Astrofisico di Catania, Citt\`{a} 
Universitaria, Viale A. Doria 6, I-95125 Catania, Italy}


\altaffiltext{1}{Also: Theoretical Astrophysics Center, Copenhagen,
Denmark} 


\begin{abstract}
Evidence for clustering within the Coma cluster is found 
by means of a multiscale analysis of the combined angular-redshift 
distribution.
We have compiled a catalogue of 798 galaxy redshifts
from published surveys from the region of the Coma cluster.
We examine the presence of substructure and of voids 
at different
scales ranging from $\sim 1$ to $\sim 16 \, h^{-1}$ Mpc, using 
subsamples of the catalogue,  ranging from $cz=3000$ km/s
to $cz=28000$ km/s.
Our substructure detection method is based on the wavelet transform and on the
segmentation analysis. The wavelet transform allows us to find out structures
at different scales and the segmentation method allows us a quantitative
statistical and morphological analysis of the sample.
From the whole catalogue we select a subset of $320$ galaxies, with 
redshifts between $cz=5858$ km/s and $cz=8168$ km/s that
we identify as belonging to the central region of Coma and on 
which we have performed a deeper analysis, on scales ranging 
from $180\, h^{-1} $ kpc to $1.44 \, h^{-1}$ Mpc.
Our results are expressed in terms of the number of structures or voids
and their sphericity for different values of the threshold detection 
and at all the scales investigated. 
According to our analysis, there is strong evidence for multiple hierarchical
substructure, on scales ranging from a few hundreds of kpc to about 
$4 \, h^{-1} $ Mpc. The morphology of these substructures is rather spherical. 
On the scale of $720\, h^{-1} $ kpc we find two main subclusters 
which where also found before, but our wavelet analysis shows 
even more substructures, whose redshift position is approximatively marked by
these bright galaxies: 
NGC 4934 \& 4840, 4889, 4898 \& 4864, 4874 \& 4839, 4927, 4875.

\end{abstract}


 

%

\section{Introduction}
The Coma cluster (number 1656 in the Abell [1958] catalogue) 
has been perhaps the most studied galaxy cluster 
since $1933$, when Zwicky calculated its mass (Zwicky, 1933).
It has been long quoted as the paradigmatic example of a 
roughly spherical, relaxed cluster (Sarazin, 1986). 
Previous papers (e.g. Fitchett \& Webster, 1987;
Mellier et al., 1988;
Baier et al., 1990; Briel et al., 1992; White et al., 1993;
Colless \& Dunn, 1996, hereafter CD96 and Biviano et al., 1996)
have suggested that this cluster may 
have a complex structure. The X-ray images obtained with 
{\sl ROSAT} suggest the presence of clumps of emission associated with 
substructures (Briel et al., 1992; White et al., 1993).   
However, previous analysis were performed only on
2-D slices of the cluster. In this paper we take up the issue of 
substructure in this cluster by yet another point of view, 
namely by trying to make use of redshift information.

Our aim is to find out substructure or voids at different scales 
through a three-dimensional analysis of the cluster, identify 
them and make a
morphological analysis. 
As it will be evident, our wavelet analysis detects substructure
which is not visible in 2-D images, either optical and/or X-ray.

The plan of the paper is as follows: in \S 2 we discuss which selection criteria
we have adopted to assemble our catalogue. In \S 3 and in \S 4 
we discuss the method of analysis based
on wavelet transform and segmentation. In \S 5 we report our results
concerning the number and morphology of substructures and in 
\S 6 we do the same for the central region of the cluster. In \S 7 we 
make some cautionary remarks concerning the statistical and physical 
significance of our analysis, and finally in \S 8 we report our conclusions.

\section{The Data}
A large body of data on the Coma cluster is available in the literature.
The catalogue has been collected exploting data coming from different 
redshift surveys. In total we have selected 798 redshift for galaxies lying 
in the range:
\begin{eqnarray} 
RA & =& 10^h  42^m  00^s,    15^h  28^m  00^s \nonumber \\
DEC & = & +26^o  30'  00",   +29^o  55'  00" \\ 
\nonumber
\end{eqnarray}
(B1950.0, hereafter all coordinates are referred to1950.0). 
The redshifts for 243 galaxies have been kindly provided to us in electronic
form by J. Colless and come from the new redshifts 
survey by CD96. The mean redshift uncertainty is: of about $50$ km/s, 
and the uncertainty in the positions is less than 1".
Another sample of 305 redshifts have been taken from
Biviano et al. (1996); 225 of them are new measurements made 
at the Canada-France-Hawaii Telescope; the 
mean uncertainty is about $100$ km/s. 
The positions in this catalogue are known with
a mean error of about 2".
Another 320 redshifts have been taken from
the catalogue of 379 galaxies by I. D. Karachentsev $\&$ A. I. Kopylov 
(1990).
The mean uncertainty in their
measurements
is  $100$ km/s, while the mean error in the positions is $\pm$ 3".
Finally, another 46 redshifts have been taken from the NED (NASA 
Extragalattic Database). This latter set of data is
heterogeneous, however their 
quoted mean uncertainty is less then $\sim 120$ km/s whilst the position are 
known with a mean error of about 6".

The total number of galaxies so collected is 914, but  
some objects are common to the three data sets.
We have considered a galaxy common when its position in 
a given data set is inside the mean error in the coordinate determination
of a different set.
So, if for the same galaxy
several redshift measurements were available,
we have included only the most accurate one.\\

\begin{table}[ht] 
\centering
\begin{tabular}{c c c c c c} 
\hline
  & \# gal & $\langle cz \rangle$  & $\sigma_{cz}$ & $(cz)_{min}$ &$(cz)_{max}$ \\ \hline
$C_{ext}$  &  690 & $10541 \pm 115$ & $6056 \pm 98$ & 3000 & 28000 \\ 
$C$  &  485 & $7013 \pm 100$ & $1155 \pm 80$ & 4000 & 10000 \\
$C_{cen}$  &  320 & $7034 \pm 80$ & $597 \pm 70$ & 5855 & 8168 \\ \hline
\end{tabular}
\caption[h]{Catalogs used in our analysis. Redshifts are in km/s}
 \end{table}

The completeness of our heterogeneous catalogue is about $85 \%$ at 
$m_B=18$. This value has been calculated from a weighted mean of 
the values for the different database used in our compilation.

To summarize, we have a heterogeneous sample of redshifts for 
798 galaxies with $1000 \le cz \le 115000$ km/s.
As coordinates for cluster's photometric centre we choose the value
quoted by Godwin et al. (1983):  
${\sc RA}=12^h 57^m.3 $ and ${\sc DEC}= +28^o 14.4'$.
The uncertainty on the redshifts measurement is about 100 km/s and 
the maximum error on the positions is less than 3".
From this catalogue we extract three different subsamples as shown in Table 1.
The line-of-sight distribution for the galaxies of our subsample with
3000 km/s $< cz <$ 28000 km/s is shown in fig. 1a.
Hereafter we call this region $C_{ext}$. 
The number of galaxies inside this region is 690. 
already this histogram seems to suggest the presence of two main peaks 
in the redshift distribution.

The subsample $C_{ext}$  has a mean redshift $\langle cz \rangle = (10541  
\pm 115)$ km/s 
and a standard deviation ${\sigma}_{cz}= (6056 \pm  98)$ km/s.
The galaxies of the first peak have redshifts $4000 \le cz 
\le 10000$ km/s;  the mean redshift and the standard deviation 
being respectively $\langle cz \rangle = (7013 
\pm  100)$ km/s and  
$\sigma_{cz} = (1155 \pm 80)$ km/s. The line-of-sight
velocity dispersion for the galaxies of the first peak is
$\sigma_{v} = (1128 \pm 78) $ km/s. 
The galaxies of the second peak  have 
redshifts $16100 \le cz \le 24600$ km/s, 
mean redshift and standard deviation 
being respectively  $\langle cz \rangle = (19918 \pm 100)$ km/s 
and  $\sigma_{cz} = (3745 \pm  80) $ km/s.
In Fig. 1b  only those galaxies inside the first peak are considered
and the  line-of-sight distribution is shown. Hereafter 
we call this region $C$.  The number of galaxies inside $C$ is 485.  
From Fig. 1b we note that the distribution of Coma galaxies
in the redshift field show the presence of some peaks and this also
may suggest the existence of multiple substructure.

We have transformed the angular distances from the
photometric center to linear one assuming a distance from the
center of the cluster equal to the mean value of the redshift
divided by $H_0$. Finally, to make our linear coordinates 
independent from the value of $H_0$, we renormalize them dividing by 
${\sigma}_{cz}$.

\section{The method of analysis: wavelet transform and segmentation.}
\subsection{The wavelet transform}
Our method of structure detection is based on the wavelet
transform evaluated at several scales
and on the segmentation analysis, and is similar to the one 
developed by Lega (1994, hereafter L94; part of this Ph.D. thesis may 
be found in Lega et al., 1995). 

A detailed description of the implementation of the algorithms is beyond the
purpose of this paper. A parallel version for a Connection Machine {\sl CM200}
has been developed by Lega (L94), a new {\sl PVM} version will be described in
Pagliaro \& Becciani (1997). 
Although the method we implement is three-dimensional for simplicity we 
describe here the one dimensional version: the generalization to the 
3-D case is straightforward.

For a one-dimensional function $f(x)$ the wavelet transform is a 
linear operator that can be
written as:
\begin{eqnarray}
w(s,t) & = & \langle f | \psi \rangle \nonumber \\
& = & s^{-1/2} \int^{+\infty}_{-\infty} f(x) \psi^* \left( \frac{x-t}{s} \right)
dx
\end{eqnarray}
where $s( > 0) $ is the scale on which the analysis is performed, $t \in \Re$ is 
the spatial translation parameter and $\psi$ is the 
the Grossmann-Morlet (1984, 1987) analyzing wavelet function 
\begin{equation}
\psi_{(s,t)}(x) = s^{-1/2} \psi \left( \frac{x-t}{s} \right)
\end{equation}
that is spatially centered around $t$ and has scale $s$.
The wavelet function  $\psi_{(1,0)}(x)$  is called {\em mother 
wavelet}. 
It generates the other wavelet function $\psi_{(s,t)}(x), s>1$.
We follow L94 in the choice of the  
mother wavelet: 
\begin{equation}
\psi_{(s,t)}(x)=\phi(x)-\frac{1}{2} \phi(\frac{x}{2})
\end{equation}
where $\phi$ is the cubic centred B-spline 
function defined by:
\begin{equation}
\phi(x) = \frac {|x-2|^3 - 4|x-1|^3 + 6|x|^3 - 4|x+1|^3 +|x+2|^3} 
{12}
\end{equation}

Although our data distribution is highly anisotropic, we prefer to use an
isotropic wavelet function and to perform a scale transformation 
along the $z$ (redshift) axis.

With these choices, the wavelet coefficients at different scales can be
calculated by the {\it \`a trous} algorithm, as described by L94 (pag.100), which is
extremely fast and requires the set of scales to be powers of two:  $s=2^r$.

The scale $s$ in this kind of analysis may be considered the resolution.
In other words, if we perform a calculation on a scale $s_0$, we expect the
wavelet transform to be sensitive to structures with typical size of about $s_0$
and to find out those structures.

\subsection{Choice of threshold}
The thresholding is made on the wavelet coefficient histogram. 
For a flat background, the wavelet transform yields zero coefficients.  
The existence of structures
at a given scale gives wavelet coefficient with large positive values. 
Obviously, a random distribution may result in coefficients different 
even if there is no structure, due to statistical fluctuations. Moreover, 
the statistical
behaviour of the wavelet coefficient is complex because of the
existence of correlation among nearby background structures which
reflect in correlations among nearby pixels.\\
In order to single out significant structures we have to fix a
tresholding criterion and level. We choose the threshold using a classical 
decision rule. We calculate the
wavelet coefficients $w_{ran}(s)$ for each scale of our analysis, 
for $10$ random distribution in the same  region of space of our data and on
the same grid.
Then we calculate the probability 
$P[w(s) \le w_{ran}(s)]$  and choose the value $w_{thres}(s)$ so that:
\begin{equation}
P[w_{thres}(s) \le w_{ran}(s)]  \le  \epsilon
\end{equation}
Our threshold on the scale $s$ is the value $\nu_{thres} = w_{thres}(s)$. 
Our choice for the value of $\epsilon$ is: 
\begin{equation}
\epsilon=0.001
\end{equation}
that ensures a $99.9 \%$ confidence level in the structure detection.

However, for the sake of completeness,
we perform our analysis for several values of the threshold, calculated in terms
of the standard deviation in the wavelet coefficient distribution of our data.

\subsection{Structure numbering through segmentation}
The second step of our analysis is the determination of connected pixels
over a fixed threshold ({\it segmentation}, Rosenfeld 1969), 
the numbering of the selected structures and
their morphological analysis. 

The segmentation and numbering consists of the analysis of the 
wavelet coefficients' matrix;  all the pixels 
associated with a wavelet coefficient greater
than the selected threshold are labeled with an integer number. 
All other pixels labels are set equal to zero. 
Then, the same label is associated with all the pixels connected in a single 
structure, in a sequential
way. So, the first structure individuated has the label '1'  and so on.
Then, for each structure we calculate volume and surface and
from them a morphological parameter. 

The detailed description of the algorithm is beyond the purpose of this
paper. However it can be described in brief as follows:
{\it Step 1}: all pixels with $w \ge \nu_{thres}$ are labeled. 
{\it Step 2}: the same label is associated with those pixels labeled 
and connected. This is done in a sequential way: the first structure 
detected has the label '1', the $N$-th one has the label 'N'.  
This requires a renumbering of most pixels. 
{\it Step 3}:  volume and surface of each structure singled 
out are calculated. 

\subsection{The morphological parameter}
In order to perform a morphological analysis we have to introduce a
morphological parameter that quantifies the sphericity of the structures.
We choose the parameter:
\begin{equation}
L(s)= K(s) \frac{V^2}{S^3}
 \end{equation}
where $V$ is the volume and $S$ is the surface, as in L94, and $K(s)$ is
a parameter that depends on the scale of analysis. 
We want $L(s)$ to have the following behaviour:  
zero for very filamentary structures and $1$ for spherical ones.
This may be achieved putting $K=36 \pi$, but only for those scales not
affected by the granular nature of the analysis. We choose the value $36 \pi$
only for the scales $s=2^r$ pixels  with $r \ge 2$. 
For the smallest scales the constant $36 \pi $ is not
adequate, since we are close to the grid resolution and the  
geometry of the substructures cannot be spherical. 
So, since we want to consider as spherical a one-pixel structure, 
we adopt the values:
\begin{equation}
K(2^r)= \left\{  \begin{array}{ll}  
216 & \mbox{if $r=0,1$} \\  
36 \pi & \mbox{otherwise} 
\end{array} \right.
\end{equation}
Then,  for every detection threshold we calculate the
values:
\begin{equation}
\langle L(s) \rangle = \sum_{i=1}^{N_{obj}} \frac{L(s)}{N_{obj}}
\end{equation}
where $N_{obj}$ is the number of objects detected at scale $s$.

\section{The voids detection method}
Voids detection method is analogue to structure detection method,
as far as "greater than" is replaced by "less than".
The voids thresholding is made on the wavelet coefficient histogram too.
The presence of voids at a given scale gives wavelet coefficient 
with large negative values.
 
Our choice for the threshold is the same as in \S $3.2$.
We calculate the wavelet coefficients $w_{ran}(s)$ for each scale of our analysis, 
for $10$ random distribution in the same  region of space of our data. 
Then we calculate the probability 
$P[w(s) \ge w_{ran}(s)]$  and choose the value $w_{voids}(s)$ so that:
\begin{equation}
P[w_{voids}(s) \ge w_{ran}(s)]  \le  \epsilon
\end{equation}
with $\epsilon=0.001$, 
that ensures a $99.9 \%$ confidence level in the voids detection.
Our threshold is the value $\nu_{voids} = w_{voids}(s)$.
Obviously, in the segmentation algorithm for the voids detection the
labeled pixels are those with the wavelet coefficients: 
$w\le \nu_{voids}$.

The determination of the voids' morphological parameter values is 
analogue to the determination of structures' morphological parameter value
as described in \S3.4.
 
\section{Substructure and voids detection in the Coma cluster}
We examine two catalogues. The first is $C_{ext}$ made of $690$ 
galaxies,  as previously said, with
redshifts between $cz=3000$ km/s and $cz=28000$ km/s. 
Our grid ensures a resolution of
about $2 \, h^{-1} $ Mpc on each of the three axis.
We examinate four different scales: 
$2.02$, $4.04$, $8.08$, $16.16 \, h^{-1}$  Mpc,
for four different values of the threshold: the $99.9 \%$ confidence 
level as described
in \S 3, and then: $3 \sigma$, 
$4 \sigma$ and  $5 \sigma$, where $\sigma$ is the 
standard deviation in the wavelet 
coefficient distribution for the selected scale.
The second catalogue investigated is $C$. 
It is made of $485$ galaxies 
with redshifts between
$cz=4000$ km/s and $cz=10000$ km/s.
Our analysis grid ensures 
a resolution of about $470 \, h^{-1} $ kpc on each of the three axis.
We examinate four different scales: 
$0.47$, $0.94$, $1.88$, $3.76 \, h^{-1}$  Mpc,
for the same four different values of the threshold as before 
We show the wavelet coefficients distributions on the four scales for the 
two catalogues in Fig.$2 {\scriptstyle \div} 3$. 
We plot the value $(w - \langle w \rangle) / \sigma$, where $w$ is the
wavelet coefficient value and $\langle w \rangle$ and $\sigma$ are the mean
and the standard deviation in the wavelet coefficient distribution, versus
$Log_{10} P(w)$, where $P(w)$ is the probability associated to 
the wavelet  coefficient $w$.

\begin{table}[ht] 
\begin{tabular}{c c c c c c} 
\multicolumn{6}{l}{1. structures} \\ \hline
scale & 99.9\%  & $ 3\sigma$ &  $4\sigma$ & $5\sigma$ & voids \\ \hline
2.02  &  60         &   63 & 55 & 64   & 65\\ 
4.04   &  15     & 4 & 4 & 4 & 2\\ 
8.08   &  2   &  1   & 1  & 1  & 4 \\ 
16.16  & 2   & 1  &  2  & 2   & 0\\ \hline
\end{tabular}
\medskip
\begin{tabular}{c c c c c c} 
\multicolumn{6}{l}{2. morphology} \\ \hline
scale & 99.9\%  & $ 3\sigma$ & $ 4\sigma $ & $5\sigma$ & voids \\ \hline
2.02  &  0.93        &   0.90 & 0.91 & 0.91 & 0.84 \\ 
4.04   &  0.52     &  0.40 & 0.45 & 0.27 & 0.13\\ 
8.08   &  0.30   &  0.27  & 0.27  & 0.31  & 0.42\\ 
16.16  & 0.17   & 0.17  &  0.30  & 0.31  & - \\  \hline
\end{tabular}
\caption[h] {Number of structures and mean
morphological parameter $\langle L \rangle$
at different scales
and over different thresolds for the catalog $C_{ext}$. 
Scales are expressed in Mpc.}
\end{table}

The curves are slightly asymmetric on all the scales, 
with a small queue towards the positive values
of the coefficients, meaning presence of substructure.  
Our results are expressed in terms of number of structures at the selected 
scales (see the Tables 2 and 3).  
Considering the $99.9 \%$ confidence level as a significance level for the
structure detection, we have overwhelming evidence for substructure 
inside the Coma cluster 
on scales from $0.47$ to $4.04 \, h^{-1}$ Mpc. On smaller scales the evidence is
certainly lesser, as one can also see from the fact that on these scales the wavelet
coefficients' histograms are more symmetric.\\
In Tables 2 and 3 we show the morphological parameter  $\langle L \rangle$.
For what concerns $C$, our substructures are rather 
spherical on the first two scales. The value of
$\langle L \rangle$
is lowered till $0.2$ on the scale $1.88  \, h^{-1}$ Mpc,
meaning a much more filamentary morphology for those structures singled
out at this resolution.
The substructure of $C_{ext}$, singled out with a greater resolution
shows a spherical morphology till the scale of $2.02 \, h^{-1}$ Mpc; more
elongated shapes are found out at the bigger scales.
In both cases the diminution of $\langle L \rangle$  
and of the number of structures as a 
function of the scale indicates a hierarchical distribution.

\begin{table}[ht] 
\centering
\begin{tabular}{c c c c c c} 
\multicolumn{6}{l}{1. structures} \\ \hline
scale & 99.9\%  & $ 3\sigma$ &  $4\sigma$ & $5\sigma$ & voids \\ \hline
2.02  &  60         &   63 & 55 & 64   & 65\\ 
4.04   &  15     & 4 & 4 & 4 & 2\\ 
8.08   &  2   &  1   & 1  & 1  & 4 \\ 
16.16  & 2   & 1  &  2  & 2   & 0\\ \hline
\end{tabular}
\medskip
\begin{tabular}{c c c c c c} 
\multicolumn{6}{l}{2. morphology} \\ \hline
scale & 99.9\%  & $ 3\sigma$ & $ 4\sigma $ & $5\sigma$ & voids \\ \hline
2.02  &  0.93        &   0.90 & 0.91 & 0.91 & 0.84 \\ 
4.04   &  0.52     &  0.40 & 0.45 & 0.27 & 0.13\\ 
8.08   &  0.30   &  0.27  & 0.27  & 0.31  & 0.42\\ 
16.16  & 0.17   & 0.17  &  0.30  & 0.31  & - \\  \hline
\end{tabular}
\caption[h] {Number of structures and mean
morphological parameter $\langle L \rangle$
at different scales
and over different thresolds for the catalog $C_{ext}$. 
Scales are expressed in $\, h^{-1}$Mpc.}
\end{table}

\begin{table}[ht] 
\centering
\begin{tabular}{ c c c c c c } 
\multicolumn{6}{l}{1. structures} \\ \hline
scale & 99.9\%  &  $3\sigma$ & $ 4\sigma$ & $5\sigma$  & voids\\ \hline
0.47  &  54         & 205 & 200 & 75 & 40  \\ 
0.94   &  21     & 42 & 35 & 25 & 12\\ 
1.88   &  3   &  3   & 2  & 3  & 0\\ 
3.76  & 2   & 1  &  1  & 2   & 0\\ \hline
\end{tabular}
\medskip
\begin{tabular}{ c c c c c c} 
\multicolumn{6}{l}{2. morphology} \\ \hline
scale & 99.9\%  &  $3\sigma$ &  $4\sigma$ & $5\sigma$ & voids  \\ \hline
0.47  &  0.95        &   0.88 & 0.89 & 0.89 & 0.94 \\ 
0.94   &  0.85     &  0.64 & 0.62 & 0.58 & 0.60 \\ 
1.88   &  0.19   &  0.42  & 0.32  & 0.60  & -\\ 
3.76  & 0.16   & 0.10  &  0.13  & 0.12  & -\\ \hline
\end{tabular}
\caption[h] {Number of structures and mean
morphological parameter $\langle L \rangle$
at different scales
and over different thresolds for the catalog $C$. 
Scales are expressed in $\, h^{-1}$ Mpc.}
\end{table}

\section{The central region of Coma}

\subsection{Substructures}
We consider a galaxy belonging to the central region of Coma if its redshift
is inside $\pm 1 \sigma$ from $\langle cz \rangle =7013$ km/s, 
where $\langle 
cz \rangle$ and
$\sigma$ are the mean and the
standard deviation in the redshifts distribution calculated on the 485 galaxies
considered in $\S 5$.
Our catalogue $C_{cen}$ is made of $320$ galaxies, with redshifts between 
$cz=5858$ km/s and $cz=8168 $ km/s. The mean redshift is 
$\langle cz \rangle = (7034 \pm 80)$ km/s and the standard deviation
$\sigma_{cz} = (597 \pm 70)$ km/s.
Our grid ensures a resolution of
about $180 \, h^{-1 }$  kpc on each of the three axis.
We examinate four different scales: $180$, $360$, $720$ and $1440 \,
h^{-1}$ kpc, 
for the usual values of the threshold ranging from $3 \sigma$ to 
$5 \sigma$ plus the $99.9 \%$ confidence level threshold,
where  $\sigma$ is the standard deviation in the wavelet 
coefficient distribution for the selected scale.
We show the wavelet coefficients distributions on the four scales in Fig.4.
These are slightly asymmetric too, with the usual small queue 
towards the positive values, meaning presence of substructure also inside the
central region of the cluster.  
Our results are expressed in terms of number of structures at the selected 
scales.  Considering the $99.9 \%$ confidence level threshold  
as significance level for the
structure detection, we have overwhelming evidence for substructure 
inside the central region of the Coma cluster 
on the first three scales investigated: $180$, $360$ and $720\, h^{-1}$ kpc 
(see the Table 4A). 
The morphological parameter  is shown in Table 4B.
Our substructures are rather spherical on all the scales but the last one, with
a value of about $0.5$ of the morphological parameter, 
shows that the shape becomes more elongated on a scale
of typical size $720  \, h^{-1}$ kpc inside the central region.

\subsection{Search for segregation}
Having identified the substructures we tried to search for any
evidence of segregation, in luminosity and/or colour. Recently 
some evidence of morphological segregation within Coma has 
been found (Andreon, 1996). Unfortunately we had not enough 
morphological information to attempt an analysis of
morphological segregation among the substructures we found.
In fig. 6 we show that there is no evidence that the different
subgroups observed within the central region of Coma differ as
far as colour distribution $b-r$ is concerned. One must however 
keep in mind that this colour index is not strongly correlated 
with absolute $b$-magnitude, so that from this figure one
cannot draw any conclusion about the presence of morphological 
segregation. We will examine these and other aspects of
luminosity functions within Coma in a forthcoming paper.

\section{Statistical Robustness and Physical Significance}
Until now we have not tried to draw from this wavelet analysis 
of the combined angular-redshift distribution any conclusion 
about the real phase- and configuration-space structure of Coma.
Before performing this further step one should verify that our
catalogue does not suffer from any systematic selection biases or
from other types of systematic effects like those induced by
redshift distortions, as described by
Reg\"{o}s \& Geller (1989) and Praton \& Schneider (1994). About
these latter we notice that they have little significance for a
cluster like Coma, because it lies at a distance of about 68  
$h^{-1}$ Mpc  and from Table 5 we notice that the velocity
dispersion of the structures found at a scale of 0.72 $h^{-1}$ Mpc are
at most of the order of 100 km/s, so the Hubble flow term is
dominant over the peculiar velocity {\em within} these
structures.\\
On one hand, one can reasonably argue that because the structures
we find are well within the nonlinear virialized region
on these scales we are probing a region of the pahse space
detached from the Hubble flow, where the linearity between 
redshift and distance is
completely lost. On the other hand one also expects that the 
phase-space distribution within the nonlinear region 
should be enough well-mixed within {\rm each} clump (if there are
any) that the substructures detected correspond to substructures
in velocity space.\\
In order to check this latter hypothesis, following a suggestion
of the anonymous referee, we have repeated the wavelet analysis
on each of 20 realizations obtained by randomly ``reshuffling'' 
the original catalogue, i.e. redistributing randomly the
redshifts among the galaxies while keeping the angular
coordinates fixed. The results are reported in Tables 6-7, and
are consistent with those found by Escalera \& Mazure (1992)
who performed a similar analysis for 2-D catalogues. The average
values of the number of structures is always smaller than
the one found in the original catalogue, showing that the
catalogue itself is probably contaminated by some uncertainty,
probably connected to the arbitrariness in the choice of the
redshift limits, by some background contaminants, etc. However,
notice for instance that at the scale 0.44 $h^{-1}$ Mpc the number of
structures found is 54 in the main catalogue, i.e. a value
$3.13\sigma$ larger than the mean given by reshuffling over the
galaxies in Table 7. This corresponds to a confidence level of
99.82\%,  i.e. a 0.18\% probability of false detection. On the
scale 720 $h^{-1}$ kpc these figures become 99.33\% for the confidence
level and 0.67\% for the probability of false detection.
Interestingly enough, the mean value of structures found on this 
scale is 5, and a closer inspection reveals that the structures
which do not disappear during the reshuffling are those numbered
from 2 to 5 in Table 5. \\
This test strengthens our confidence on the physical
significance of most of the structures detected, particularly
when filtering on the scale of 720 $h^{-1}$ kpc. Under this respect our 
results are consistent with those found by Escalera \& Mazure 
(1992) on 2-D maps of simulated clusters, which demonstrated the
ability of the wavelet analysis to recover substructures which
are traced even by few objects. We will perform a more quantitative
analysis of the statistical significance of our wavelet analysis
of combined angular-redshift catalogues in a forthcoming paper
(Pagliaro et al., 1997a).

\section{Conclusions and discussion}
During the last years new redhift surveys and methods of analysis 
have allowed a more through understanding of structure of the 
Coma cluster (see e.g. Mellier et al., 1988; Escalera et al., 1992;  CD96; 
Biviano et al., 1996), with most of the effort going to
ascertain whether it can be 
classified as a relaxed one or not and to unveil hidden 
substructures. While this cluster has often in the past been modelled under the
assumptions of homogenous velocity structure and spherical symmetry (see e.g. 
Kent \& Gunn, 1982), the most recent observational evidence is pointing toward a
more complex structure. 
The recent {\sl ROSAT} images and 2-D optical analysis 
have strengthened the evidence for the 
existence of multiple substructure and suggest that Coma can not 
be considered a relaxed cluster. Under this respect, it is worth mentioning that
already in 1988 Mellier et al. (1988), by analysing the
isopleths within a 2-D map of the cluster had suggested the
possible existence of $9$ density peaks. 
In this paper, we have investigated the nature of the Coma cluster
performing a 3-D analysis of the combined
angular-redshift distribution of the cluster.
We have assembled a catalogue of 798 galaxy redshifts, the largest presently
available for the Coma cluster. 
Then, we have developed a 3-D wavelet and
segmentation structure analysis that has allowed us to find out 
substructures on different scales and to describe them in a quantitative way.
This powerful method of analysis has already provided excellent results
in many fields of physics (e.g. Arneodo et al.,1988; Argoul et al., 1989; 
Slezak et al., 1990; Fujiwara \& Soda, 1995; Grebenev et al., 
1995).

Our results suggests that Coma can not be considered a regular
cluster of galaxies, but it is filled up with substructure on all
scales ranging from $720$ kpc to $\sim 4 \, h^{-1}$ Mpc.
The general diminution of the mean morphological parameter, meaning
more elongated shapes,  and of the number of structures 
with the scale indicates a hierarchical
distribution of the substructure.

We have examinated the Coma cluster using three different
subsamples of our catalogue (see Table 1); so we have insights within regions 
of different sizes with different resolutions. 

On a scale of about $2 \, h^{-1} $ Mpc, our analysis on the extended 
Coma catalogue suggests the presence of multiple substructure with 
spherical morphology (see Table 2). 
On this scale a large number of voids is detected and their shapes are rather
spherical.
On the same catalogue, multiple substructure is still present at the scale $4 \, 
h^{-1} $ Mpc: on this scale shapes are more elongated ($\langle L \rangle =0.67$). 
On scales 
$8 \, h^{-1} $ Mpc and $16 \, h^{-1} $ Mpc we find only two very elongated 
objects, in agreement with the histogram of Fig.1a.
Voids on scales larger than $2 \, h^{-1} $ Mpc are few and very elongated.
Although we cannot draw any conclusion before having made a comparison
with N-body simulations (Pagliaro et al., 1997b), the presence of
hierarchically organized substructure seems to point to an evolutionary scenario
in which the Coma cluster and the galaxies included in the
second peak of the histogram of the galaxy distribution in Fig. 1a
were not generated by the collapse of two large spherical density
perturbation with different masses and radius of about $15 \, h^{-1}$ Mpc,
but by the merging of a large number of isolated spherical density
perturbations of radius ranging from $1 \, h^{-1}$ Mpc to $3 \, h^{-1}$ Mpc.
This first rough picture of the Coma evolution becomes more evident  
if we examine the catalogues $C$  and $C_{ext}$.

Our analysis of the second catalogue ($C$)
suggests the presence of substructures on all the scales with shapes becoming
more elongated with growing scale (see Table 3)
Voids are detected only on scales $0.47 {\scriptstyle \div} 0.94 \, h^{-1} $ Mpc and their
shapes are rather spherical ($0.60 \le \langle L \rangle \le 0.94$)

On smaller scales (hundreds of kpc) we have concentrated our analysis
on a central region with redshifts $5858 \le cz \le 8168 $ km/s. 
This region includes the core of Coma with the galaxies NGC 4874 
($\langle cz \rangle =$ 7131 km/s) and NGC 4839 ($\langle cz \rangle =$ 
7397 km/s). 
Multiple substructures are found on scales 
$180 {\scriptstyle \div} 720 \, h^{-1} $ kpc with rather spherical morphology
($0.50 \le \langle L \rangle \le 1.00$). 
A large number of spherical ($\langle L \rangle = 0.98$)
voids is detected only on the smallest scale ($180 \, h^{-1} $ kpc). We stress
however once again the fact that the interpretation of substrctures on such small
scales in terms of {\em real}  substructures in velocity (or position) space is not as
strong, as noted in the previous paragraph.

Finally, we concentrate on the 
scale of $720 \, h^{-1} $ kpc, where we have detected seven
substructures that we show in Fig. 5  and which coincide with the peaks that 
we find in the central region of the histogram of Fig.1b.
To each one of these object we can associate a dominant galaxy.
Mean redshifts of the objects are: $cz \sim 5912$ km/s, $ cz \sim 6100$ km/s, 
$cz \sim 6421$ km/s, $ cz \sim 6775$ km/s, $cz \sim 7161$ km/s, 
$cz \sim 7594$ km/s and $cz \sim 7805 $ km/s.
In Table 5 we report some statistics only for the clumps containing a significant
number of objects.
To each clump singled out we can associate one or two dominant galaxies.
These are (for increasing redshift):
NGC 4934 \& NGC 4840, NGC 4889, NGC 4898 \& NGC 4864, NGC 4874 \&
NGC 4839, NGC 4927, NGC 4875.
We then confirm the presence of the two subclusters already described by CD96, 
but in addition we have found statistical evidence for the exstence of more 
substructures in redshift space.
All this evidence leads us to suggest that the Coma cluster can not be considered
a regular cluster of galaxies and that its process of formation
occurs through a {\sc bottom-up} mechanism as predicted by
{\sl CDM} and {\sl MDM} models.

Finally, we would like to stress the fact that from this analysis it
is difficult to draw any conclusion about the evolutionary state
of this cluster. Such an analysis would require some modelling
of the evolutionary scenarios through comparison with
high-resolution N-body simulations and a better understanding of
the velocity field around Coma. We will report on these issues
in a subsequent work (Pagliaro et al., 1997b).

\newpage
\begin{table}
\centering
\begin{tabular}{c c c c c c c } 
\multicolumn{7}{l}{Structures at 720 $\, h^{-1}$ kpc} \\ \hline
N. Clump & N. Gal. & $cz_{min}$ & $cz_{max}$ & $ <cz>$ & $\sigma_{cz}$ & $M_{vir}$ \\   \hline
1  &  18  &  7756 & 7866 & 7805 & 30 & 6.9  \\ 
2   &  34  &  7510 & 7666 & 7594 & 39 & 16.8 \\ 
3   & 71  & 6980 & 7350  & 7161 & 105 & 284 \\ 
4  & 68  & 6572 & 6967 & 6775 & 102 & 286 \\ 
5 & 22 & 6356 & 6491 & 6421 & 34 & 10.9 \\
6 & 19 & 6028 & 6164 & 6100 & 36 & 12.4 \\  \hline
\end{tabular}
\caption[h] {Substructures detected on the scale 720 $\, h^{-1}$ kpc. 
Redshifts are in km/s. First column gives the number of the clump, the second
the number of galaxies contained within the clump, the third the NGC number
of the brightest galaxy, and the fourth  the values of the masses are in unit
 of $10^{11} M_{\odot}$.}
\end{table}

\newpage

\begin{table}
\centering
\begin{tabular}{c c c c c } 
\multicolumn{5}{c}{Statistical Tests on $C_{cen}$} \\ \hline
scale & $\langle n \rangle$ & $\sigma_{n}$ & ${\rm n}_{min}$ & 
${\rm n}_{max}$ \\ \hline

0.18 & 23.29 & 1.9 & 17 & 25\\
0.36 & 11.12 & 1.57 & 9 & 15\\
0.72 & 5.05 & 0.72 & 4 & 6\\
1.44 & 1.05 & 0.23 & 1 & 2\\ \hline
\end{tabular}
\caption[h] {Statistics on a number of ``reshufflings'' 
of redshifts in catalogue $C_{cen}$. In column 1 the scale
in  $\, h^{-1}$ kpc is reported, columns 2-5 give the average,
standard deviation, minimum and maximum number of structures
found, respectively.}
\end{table}

\medskip

\begin{table}
\centering
\begin{tabular}{c c c c c } 
\multicolumn{5}{c}{Statistical Tests on $C$} \\ \hline
scale & $\langle n \rangle$ & $\sigma_{n}$ & ${\rm n}_{min}$ & 
${\rm n}_{max}$ \\ \hline

0.47 & 42.12 & 3.79 & 33 & 48\\
0.94 & 17.47 & 3.47 & 12 & 22\\ 
1.88 & 2.29 & 0.57 & 1 & 3\\
3.76 & 1 & 0 & 1 & 1\\ \hline
\end{tabular}
\caption[h] {Same as Table 6 for catalogue $C$.}
\end{table}

\acknowledgments
We would like to thank the anonymous referee for insightful comments
which led to the introduction of the section on the statistical 
significance of the analysis performed in this paper.
A.Pa. would like to thank E. Lega for having sent him her structure 
detection code, a copy of her Ph.D. dissertation and for 
kind and indispensable help during the period in which our
code was developed and tested, and A. Bijaoui for a stimulating 
discussion held in Erice.
M.Ga. wish to thank S. Shandarin for a clarifying discussion and 
helpful suggestions.

\clearpage

\figcaption[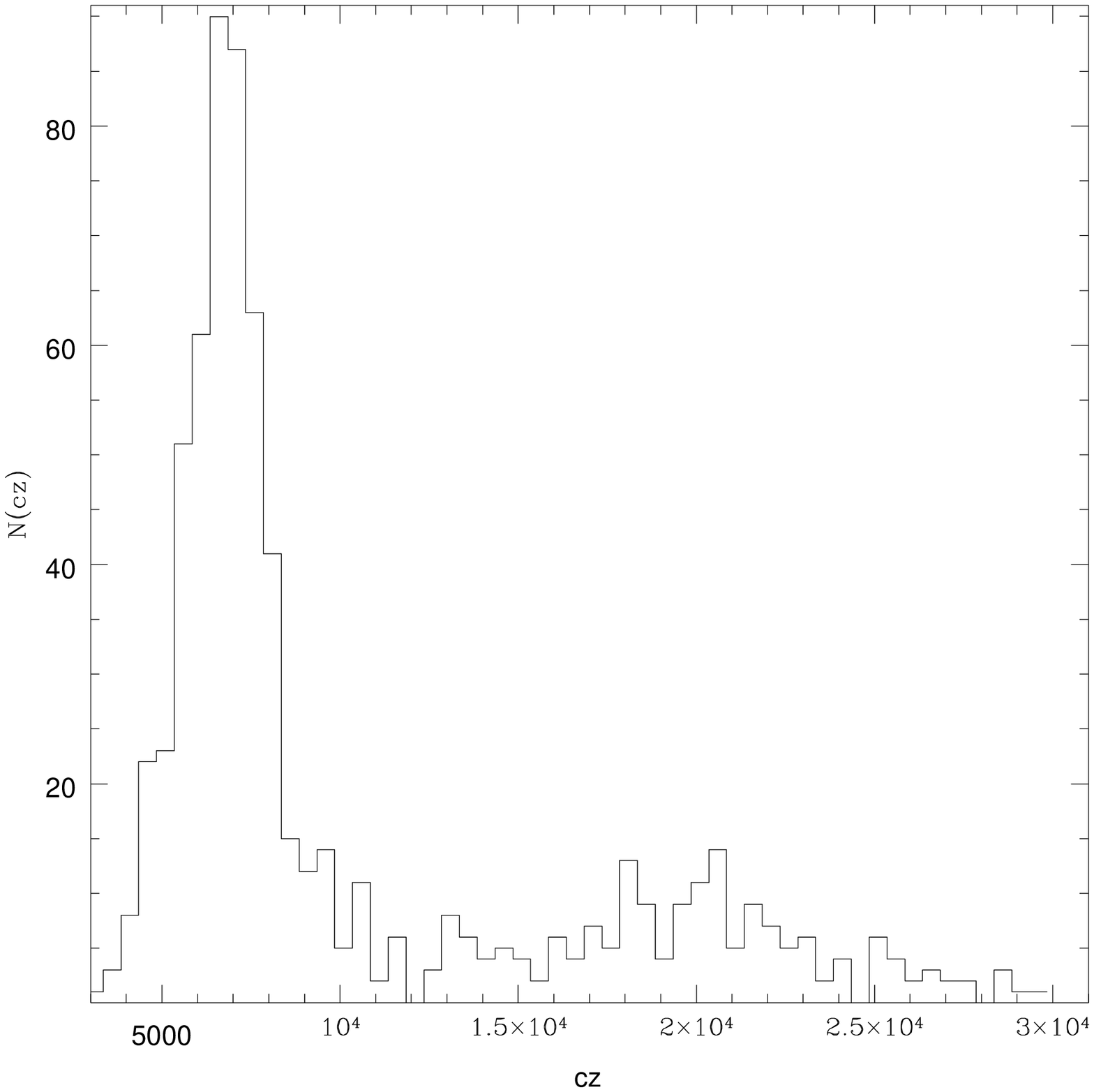, 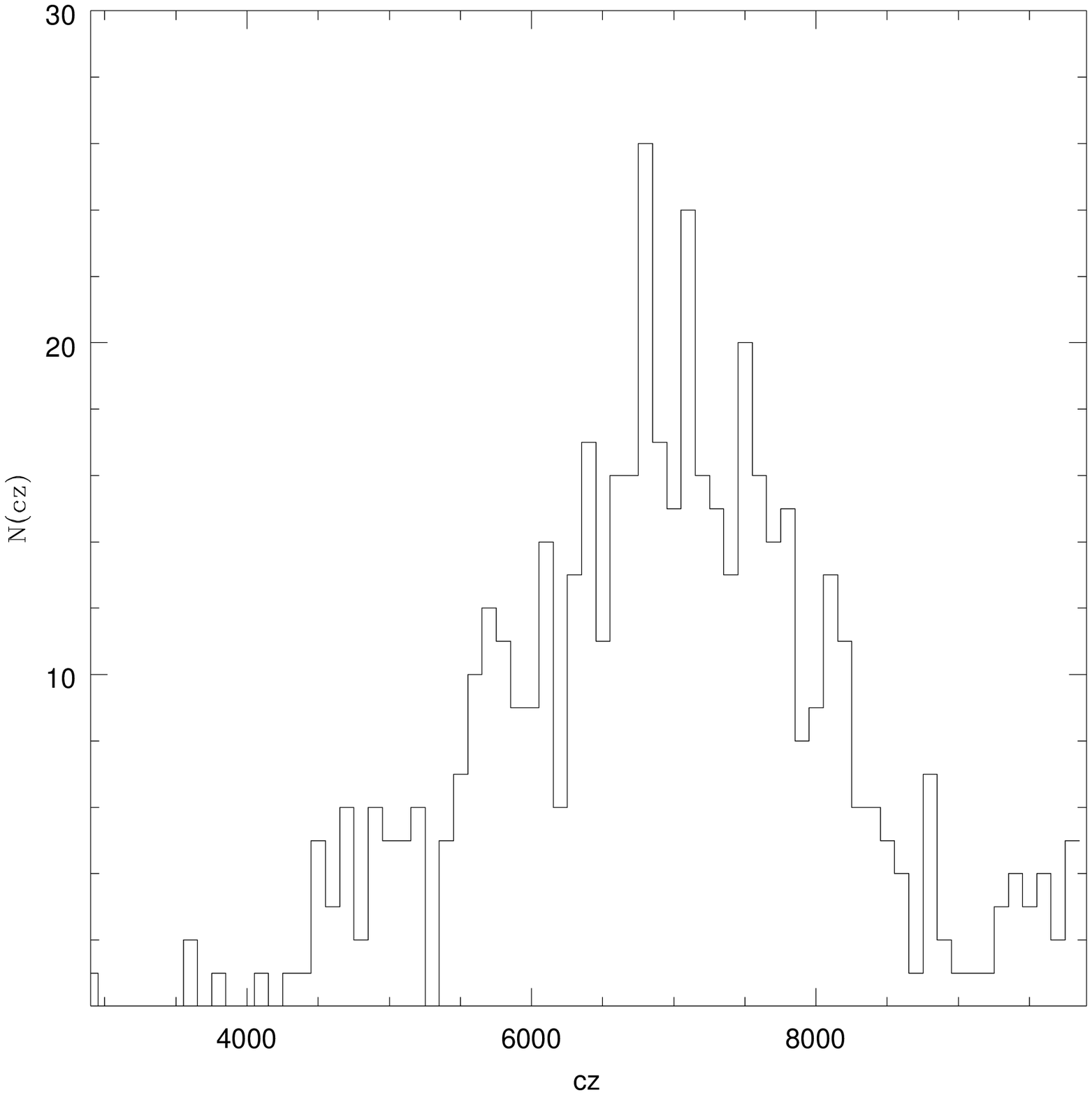]
{{\it (a)} Histogram of the galaxy distribution, with redshifts
$3000 \le cz \le 28000$ km/s inside $C_{ext}$, 
according to our catalogue. Step is 500 km/s;
{\it (b)} histogram of the galaxy distribution, with redshift
$4000 \le cz \le 10000$  inside $C$, according 
to our catalogue. Step is 100 km/s.}

\figcaption[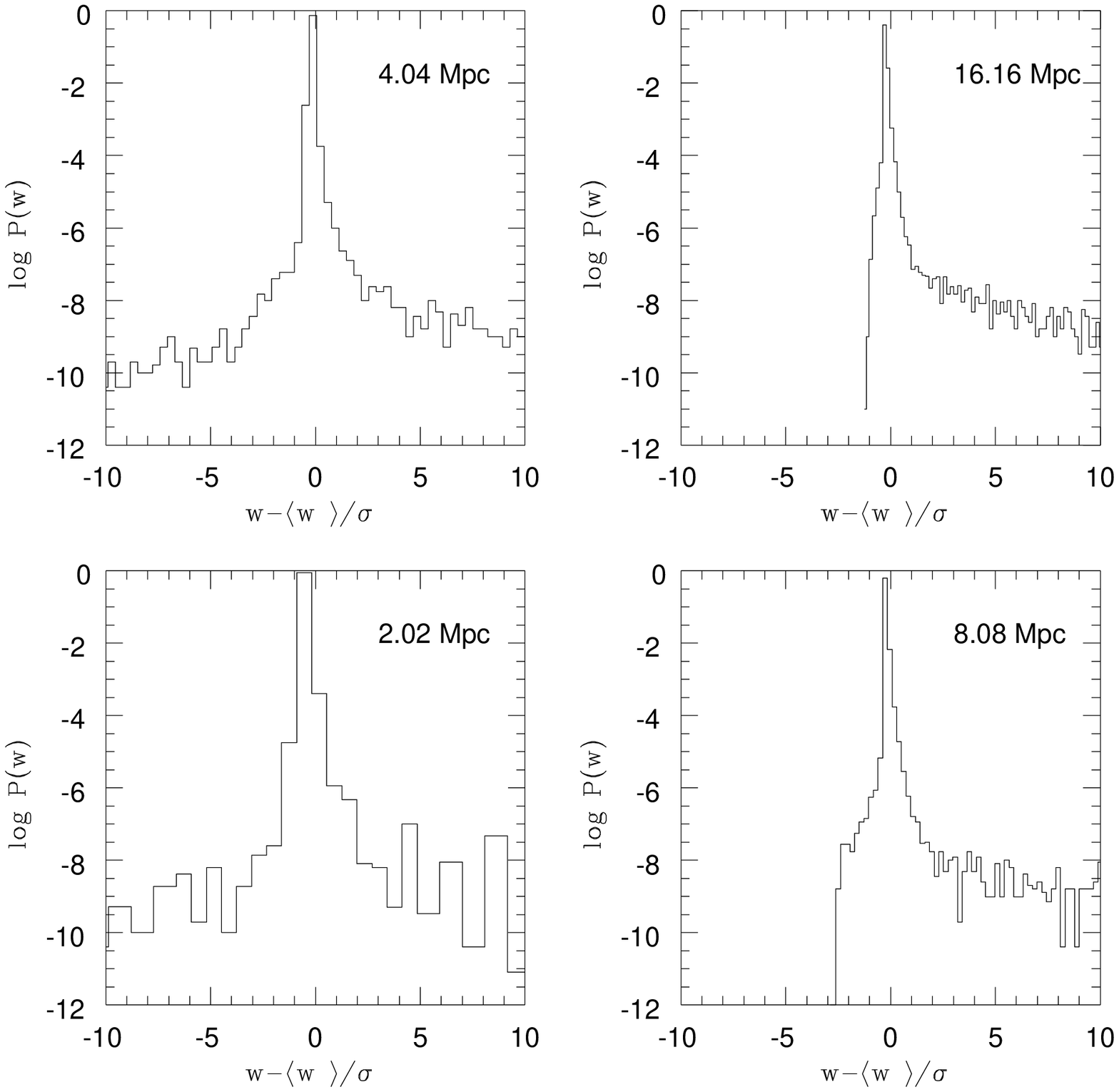]{Histograms of the wavelet coefficients on the four scales
selected for $C_{ext}$}

\figcaption[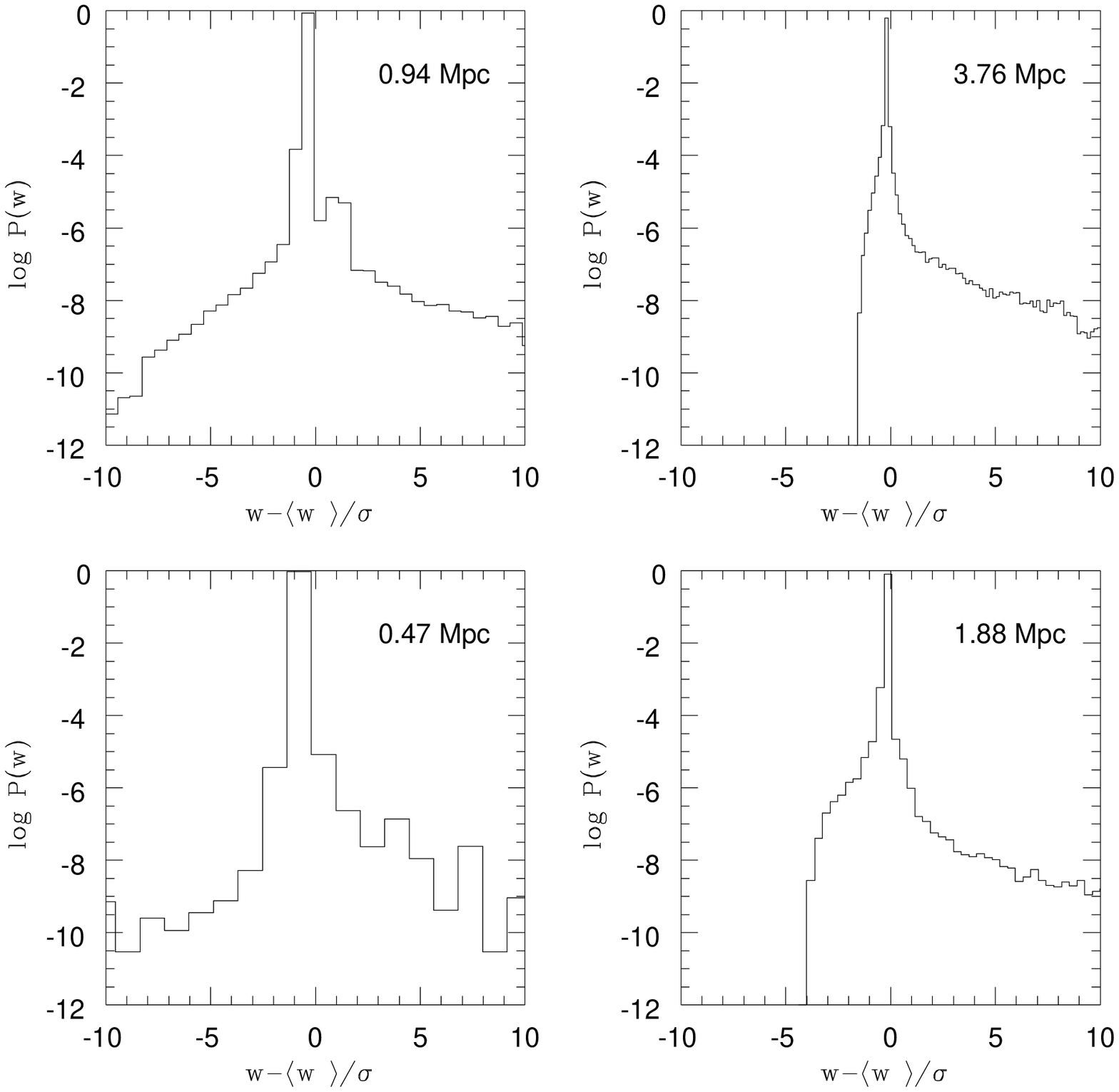]{Histograms of the wavelet coefficients on the four scales
selected for $C$}

\figcaption[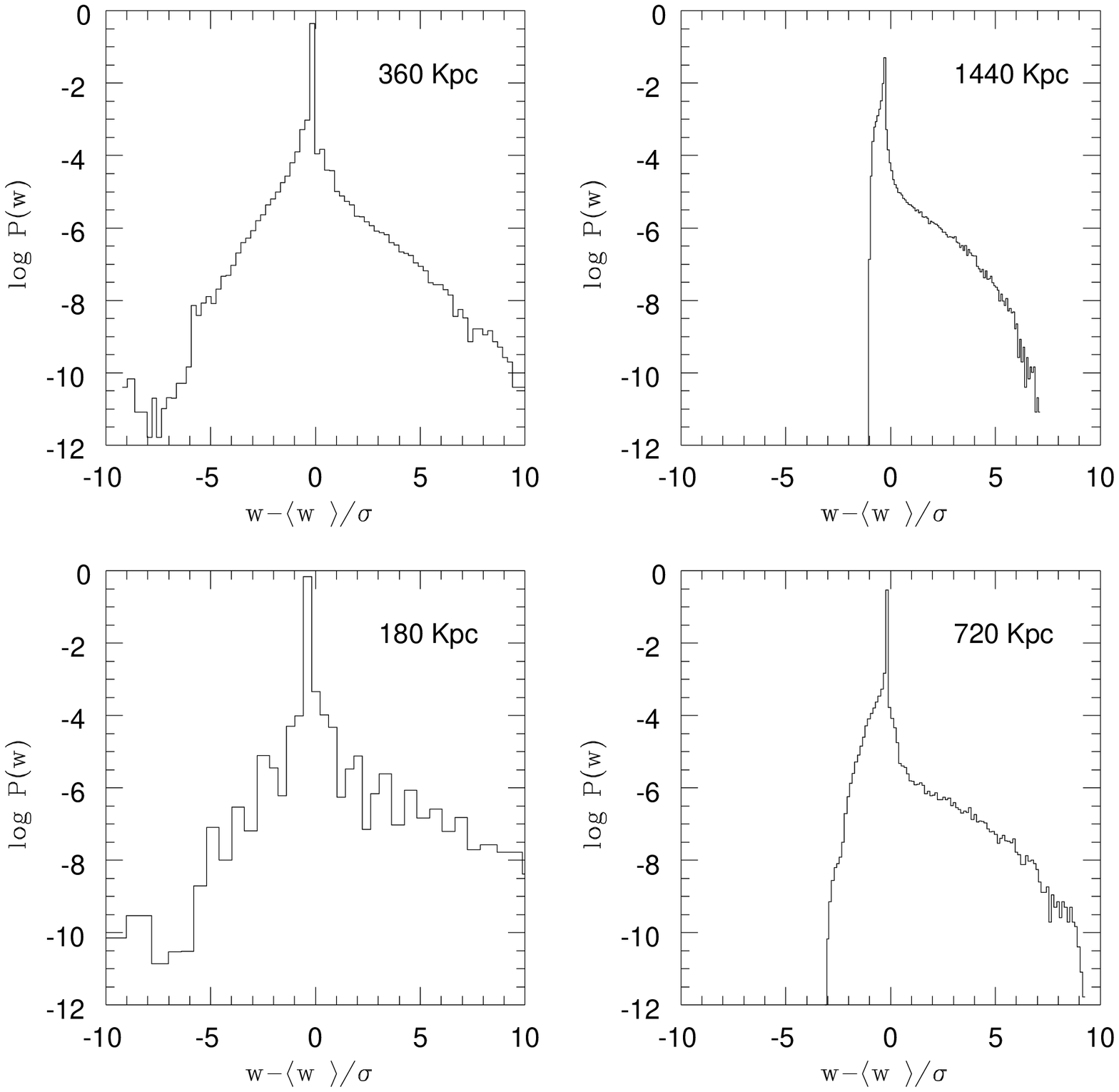]{Histograms of the wavelet coefficients on the four scales
selected for $C_{cen}$}

\figcaption[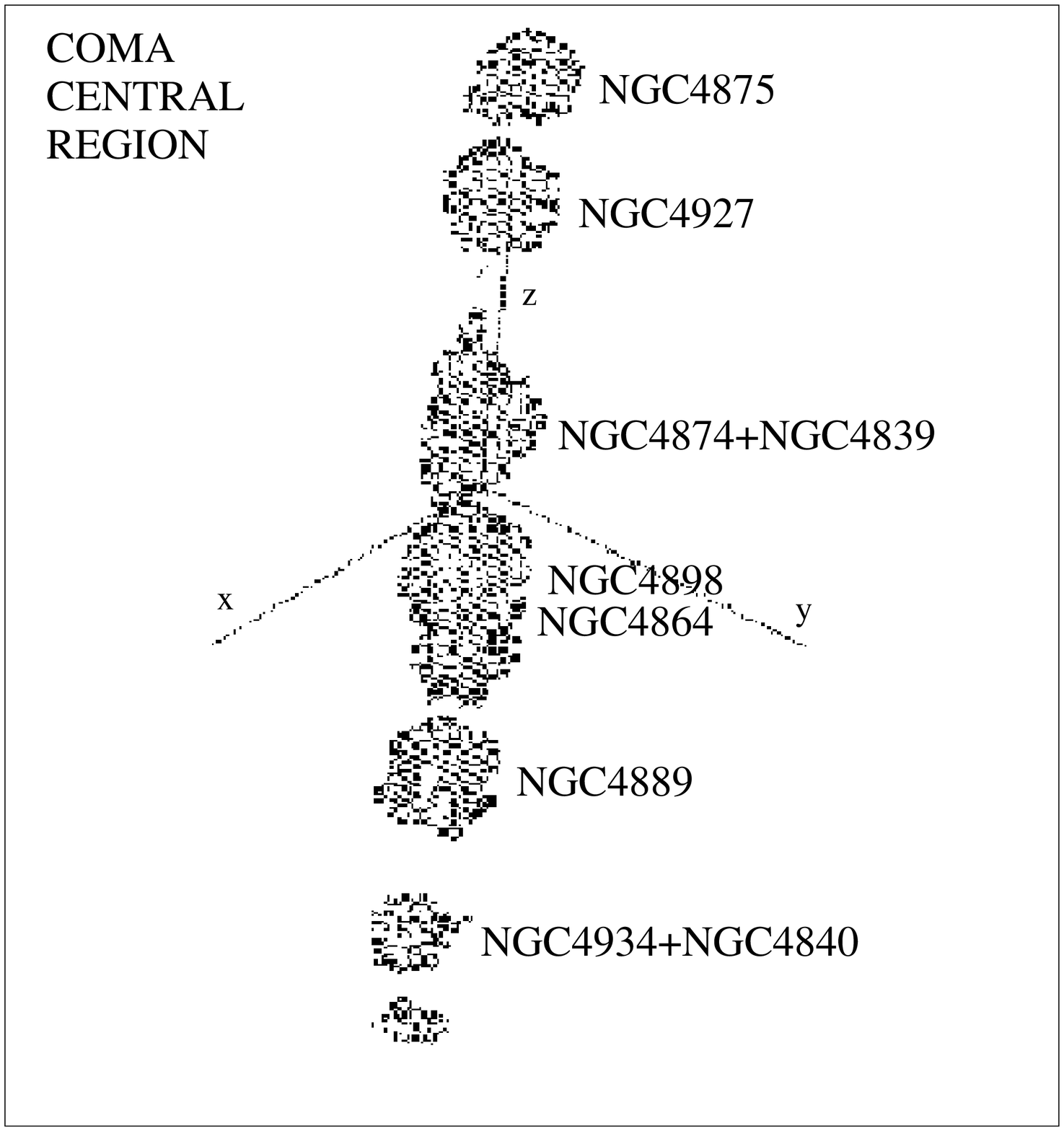]{Substructures for the central region of 
Coma detected with a resolution of
$720 \, h^{-1} $ kpc. 
The seven structures detected are clearly visible.
Morphological parameter: $\langle L \rangle = 0.50$}

\figcaption[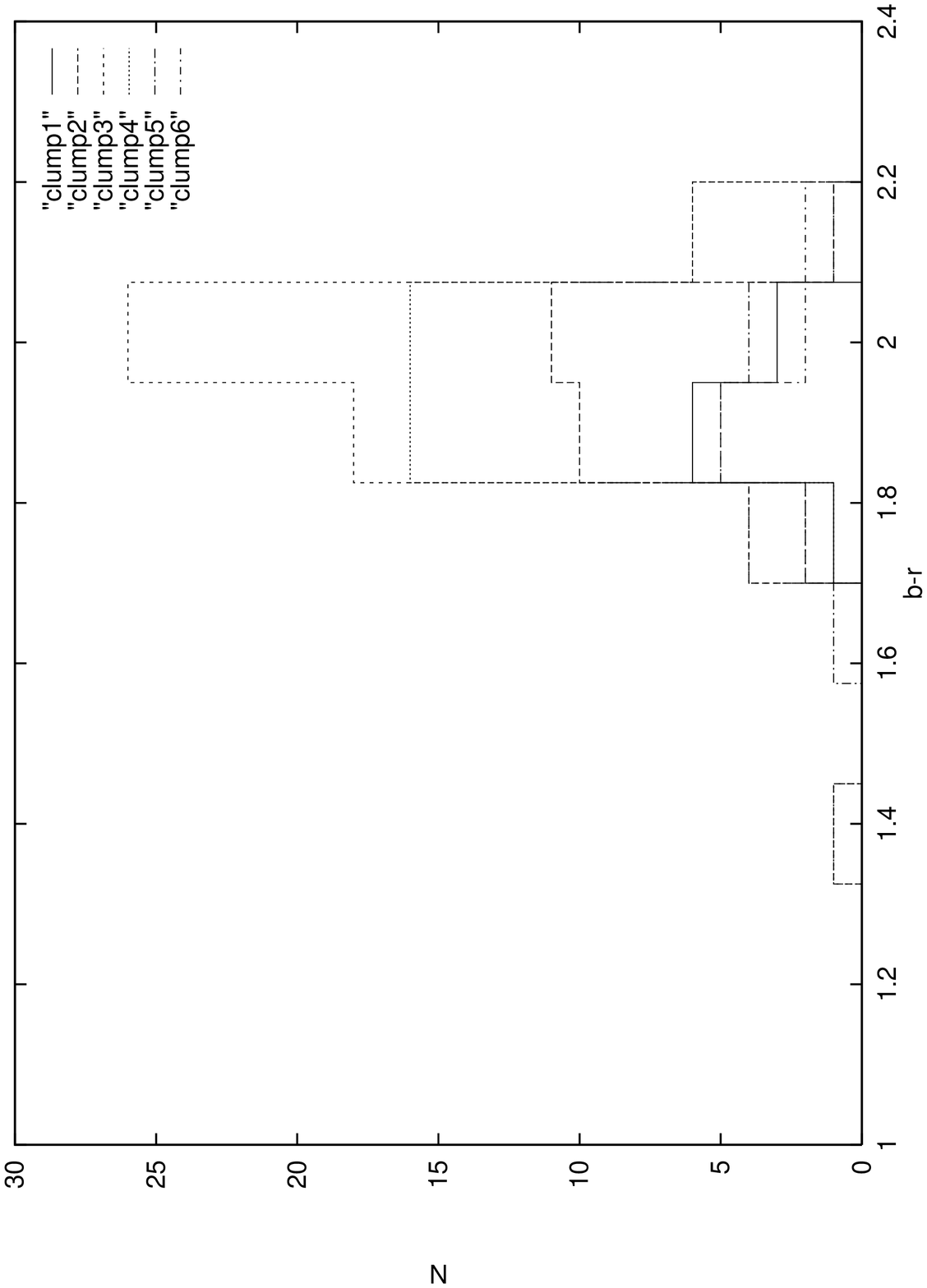]{Colour segregation. The different histograms
refer to the substructures of Table 5, and are numbered according
to the first column of that table.}


\clearpage

\begin{thebibliography}{}
\bibitem {Ab} Abell, G.O., 1958, \apjs, 3, 211
\bibitem {An} Andreon, S., 1996, \aap, 314, 763
\bibitem{Arg} Argoul, F., Arneodo, A., Grasseau, G., Gagne, Y., Hopfinger, E.F. and
Frisch, U., 1988, \nat, 338, 51
\bibitem{Arn} Arneodo, A., Grasseau, G. and Holschneider, M., 1988, \prl, 61, 2281
\bibitem{Ba} Baier, F.W., Fritze, K., Tiersch, H., 1990, Astron. Nachr., 311, 89
\bibitem{Biv1} Biviano, A., Durret, F., Gerbal ,D., Le Fevre, O., Lobo, C.,
Mazure, A., and Slezak, E., 1996, \aap Suppl. Ser., 111, 265 
\bibitem{br} Briel, U.G., Henry, J.P., B\"{o}hringer, H., 1992,\aap, 259, L31
\bibitem{col1} Colless, M., Dunn, A., M., 1996, \apj, also astro-ph/9508070 (CD96)
\bibitem{EM} Escalera, E., and Mazure, A., 1992, \apj, 388, 23
\bibitem{Esc} Escalera, E., Slezak, E., Mazure, A., 1992, \aap, 264, 379
\bibitem{FW} Fitchett, M.J., Webster, R.L., 1987, \apj, 317, 653
\bibitem {Fu} Fujiwara, Y. \& Soda, J., 1995, astro-ph/9509102
\bibitem{God1} Godwin, J., G., Metcalfe, N., Peach, J., V., 1983, \mnras 
202, 113
\bibitem{Greb} Grebenev, S.A., Forman, W, Jones, C. and Murray, S., 1995,
\apj, 445, 607 
\bibitem{GM} Grossmann, A., Morlet, J., 1984, SIAM J. Math., 15, 723
\bibitem{GM2} Grossmann, A., Morlet, J., 1987, Math. \& Phys., Lectures on
recent results, ed. L.Streit, World Scientific
\bibitem{Kar1} Karachentsev, I., D., Kopylov, A., I., 1990, \mnras, 243, 390
\bibitem{KG} Kent, S.M. and Gunn, J.E., 1982, \aj, 87, 945
\bibitem{leg1} Lega, E., 1994, These de Doctorat, Universit\'e de Nice (L94)
\bibitem{leg2} Lega, E., Bijaoui, A., Alimi, J.M., Scholl, H., 1996
\apj, also astro-ph/9510156
\bibitem{leg3} Lega, E., Scholl, H., Alimi, J.-M., Bijaoui, A., Bury, P., 1995,
Parallel Computing, 21, 265
\bibitem{Maz} Mazure, A., Proust, D., Mathez, G., Mellier, Y., 
1988, \aap, 75, 339
\bibitem{Mell} Mellier, Y., Mathez, G., Mazure, A., Chauvineau, B. and Proust, D., 
1988, \aap, 199, 67
\bibitem {psc} Praton, E.A. and Schneider, S.E., 1994, \apj, 422, 46
\bibitem {PBGA} Pagliaro, A. and Becciani, 1997,  A parallel code for structure detection (in preparation)
\bibitem {PGA} Pagliaro, A., Antonuccio-Delogu, V., Becciani, U., Gambera, M., 1997a, (in preparation)
\bibitem {PGA2} Pagliaro, A., Antonuccio-Delogu, V., Becciani, U., Gambera, M., 1997b, (in preparation)
\bibitem {rg} Reg\"{o}s, E. and Geller, M.J., \aj, 98, 755
\bibitem{rose} Rosenfeld, A., 1969, Picture Processing by Computer, Academic
Press, New York 
\bibitem{Sle} Slezak, E., Bijaoui, A., Mars, G., 1990, \aap, 227, 301
\bibitem{Whi} White, S.D.M., Briel, U.G., Henry, J.P., 1993, \mnras,  261, L8
\bibitem{zwi} Zwicki, F.,1933, Helv.Phys.Acta 6, 10
\end{thebibliography}
\end{document}